\documentclass[twocolumn,showpacs,preprintnumbers,amsmath]{revtex4}
\usepackage{amssymb}

\usepackage{graphicx}
\usepackage{dcolumn}
\usepackage{bm}

\begin{document}

\title{Nernst effect in the two-dimensional XY model}
\author{Qing-Hu Chen}
\address{Center for Statistical and Theoretical Condensed Matter Physics,
Zhejiang Normal University, Jinhua 321004, P. R. China \\
and Department of Physics, Zhejiang University, Hangzhou 310027, P.
R. China}

\date{\today}

\begin{abstract}
We calculate the Nernst signal  directly in the phenomenological
two-dimensional XY model.  The obtained numerical results are
consistent with the experimental observations in some high-Tc
cuprate superconductors qualitatively, where the vortex Nernst
signal has a characteristic tilt-hill profile. It is suggested that
the excitations of vortex and anti-vortex in 2D is the possible
origin of the anomalous Nernst effect.
\end{abstract}


\maketitle

By a phase only model, Podolsky et al\cite{Podolsky} have studied
the Nernst effect \cite{xu} and diamagnetic response due to the
thermal superconducting fluctuation through the transverse
thermoelectric conductivity ($\alpha_{xy}$). They found that the
Nernst effect has a much sharper temperature decay than predicted by
Gaussian fluctuations.  We think that for this phase model the
Nernst signal, $e_N=\alpha_{xy}/\sigma$ ($\sigma$ is the electrical
conductivity), can be calculated directly and can be compared
directly with the experiments. For some cuprate superconductors,
such as single-layer $Bi_2Sr_{2-y}La_yCuO_6$,  the quasi-particle
contribution to the Nernst signal is virtually negligible for $y$ in
the range $0.4-0.6.$ \cite{Ong}. The phase model can only be applied
to such kind of  cuprate superconductors. Note that
$\alpha_{xy}=\alpha^{S}_{xy}+\alpha^{N}_{xy}$, which includes the
contributions from both vortices  and quasi-particle, and can not be
measured directly in experiments. It is also difficult to separate
the contributions from vortices and quasi-particles.  The
contribution from quasi-particles $\alpha^{N}_{xy}$ can not be
obtained in this model. Therefore the $\alpha_{xy}$ in Ref.
\cite{Podolsky} in principle can only describe the vortex dominated
Nernst effect, and  can not describe the Nernst effect in those
cuprate superconductors where the quasi-particle signal is
comparable with the vortex one.

The same Hamiltonian of the two dimensional (2D) XY  model is given
by\cite{Podolsky} $ H=-J\sum_{\langle ij\rangle }\cos (\phi
_{i}-\phi _{j}-A_{i,j})$, where  $A_{ij}$ represents the magnetic
vector potential of a field $\mathbf{B} =\nabla \times \mathbf{A}$
perpendicular to $xy$ plane. The average number of vortex lines per
plaquette $f=a^{2}B/\Phi _{0}$, with $a$ grid spacing. With the
model-A dynamics,  the equations of motion are given by $ \hbar
\frac{d\phi _i(t)}{dt}=-\Gamma \frac{\partial H}{\partial \phi _i}
+\Gamma _i(t)$ where $\Gamma $ is a dimensionless constant which
determines the time scale
of relaxation. The thermal noise term $\Gamma _i(t)$ is assumed to satisfy $%
\left\langle \Gamma _i(t)\right\rangle =0$ and $\left\langle \Gamma
_i(t)\Gamma _j(0)\right\rangle =2\hbar \Gamma k_BT\delta _{ij}\delta
(t)$. The time and the temperature are in units of $\hbar/\Gamma J$
and $J/k_B$. The present simulations are performed with the system
size $L=64$ for two directions.

The temperature gradient is set up along x direction, where a open
boundary condition should be applied.  Specifically, we introduces a
global "twist" variable $\Delta_y(t)$ to track the average phase
drop per link in the $y$-direction \cite{kim,chen}. The voltage by
the diffusion of free vortices thermally driven is given by  $
v_y=\dot{\Delta}_y$. Since the periodic boundary condition for the
phase variable is satisfied in the $y$ -direction, previous
pseudo-spectral algorithm \cite{chen} can also be employed to
facilitate the simulations. The time stepping is done using a
second-order Runge-Kutta scheme with $\Delta t=0.05$. The
measurement is performed after the steady-state is reached. It is
found that the present thermally driven system is time-consuming
considerably. Our runs are typically $(4-8)\times 10^{9}$ time
steps. The detailed procedure in the simulations will be  described
in a full paper \cite{chen1}.

We calculate the Nernst signal directly through
$e_N=\frac{v_y}{\left| \nabla _xT\right| }$. As shown  in Fig. 1(a),
our numerical results are consistent with those in Figs. 11 and 12
in Ref. \cite{Ong} qualitatively, where the vortex Nernst signal has
a characteristic tilt-hill profile. At each temperature, in the low
magnetic field regime, the interaction between vortices is small and
can be neglected, so the Nernst signal is added linearly with the
addition of free vortices as the fields increase. In the high
magnetic field regime, the strong interaction among the dense
vortices prohibits the vortex motion driven thermally, and cause
very small Nernst signal.  The competition of these two mechanisms
results in a maximum Nernst signal in the intermediate fields. This
nonmonotonous behavior is clearly observed in the curves for the
Nernst signal  at various temperatures above and below $T_{KT}$ in
the present simulations. For fixed fields, as the temperature
increases, the Nernst signal also displays nonmonotonous behavior,
also similar to the experimental observations\cite{Ong}. It is
suggested that the excitations of vortex and anti-vortex in 2D is
the possible origin of the anomalous Nernst effect observed in
strongly layered high-Tc cuprate superconductors. The inter-layer
Josephson coupling only renormalizes the in-plane coupling strength
and may play no essential role in this phase-disordering scenario.

More interestingly, our numerical results for the Nernst signal can
explain the ridge fields in the Nernst experiments well \cite{Ong}.
In this model, the critical field $H_{c2}$ is corresponding to
$f=1$, i.e. the vortices occupy the system completely. The peak
position around $f=0.1$ shown in Fig. 1(a) is consistent with the
typical ridge fields  $0.1H_{c2}$ observed in the Nernst experiments
\cite{Ong}.

\begin{figure}[tbp]
\centering
\includegraphics[width=6.5 cm]{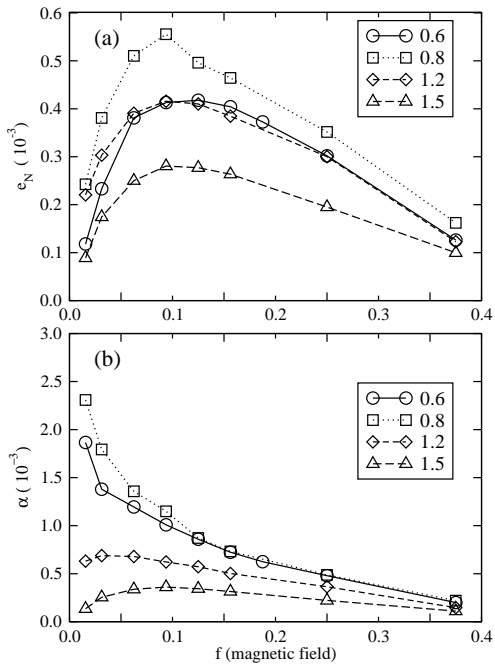}
\caption{(a) $e_N $ and (b)  $\alpha _{xy}$ as a function of $f$ at
various temperatures above and below $T_{KT}=0.89$. The system size
$L=64$.} \label{fig1}
\end{figure}

In Fig. 1 of Ref. \cite{Podolsky}, one can find that the  transverse
thermoelectric conductivity $\alpha _{xy}$  decreases monotonically
with temperatures. In our simulations, we find a nonmonotonous
behavior of $\alpha _{xy}$ below  $T_{KT}$, as shown in Fig. 1(b).
This numerical prediction for the only contribution from vortices in
$\alpha _{xy}$ is clearly  called for further confirmation in
further advanced experiments where the vortex contribution is indeed
separated from the total $\alpha _{xy}$. To the best of our
knowledge,  at the moment  no one can separate the vortex
contribution easily.

In short, the direct simulation for the Nernst signal in the
phenomenological 2D XY model can provide a qualitatively explanation
to the experimental findings in some high-Tc cuprate
superconductors.

 The author acknowledges useful discussions with Z. A. Xu
and Y. Y. Wang. This work was supported by National Natural Science
Foundation of China under Grant Nos. 10774128.

\end{document}